\documentclass[preprint,showpacs,preprintnumbers,amsmath,amssymb]{revtex4}

\usepackage{bm}

\newcommand{\de}{\delta}

\newcommand{\la}{\lambda}
\newcommand{\om}{\omega}

\newcommand{\eps}{\varepsilon}

\newcommand{\PP}{\mathcal{P}}
\newcommand{\M}{\mathcal{M}}

\newcommand{\Qb}{\bar{Q}}

\newcommand{\Ah}{\hat{A}}
\newcommand{\Bh}{\hat{B}}
\newcommand{\Ch}{\hat{C}}

\newcommand{\Fh}{\hat{F}}

\newcommand{\Hh}{\hat{H}}

\newcommand{\Kh}{\hat{K}}
\newcommand{\Nh}{\hat{N}}

\newcommand{\Qh}{\hat{Q}}

\newcommand{\Sh}{\hat{S}}
\newcommand{\Th}{\hat{T}}
\newcommand{\Uh}{\hat{U}}

\newcommand{\fh}{\hat{f}}
\newcommand{\gh}{\hat{g}}
\newcommand{\PPh}{\hat{\mathcal{P}}}
\newcommand{\lah}{\hat{\lambda}}
\newcommand{\omh}{\hat{\omega}}
\newcommand{\phih}{\hat{\phi}}
\newcommand{\Kbh}{\hat{\bar{K}}}
\newcommand{\Qbh}{\hat{\bar{Q}}}

\newcommand{\pa}{\partial}

\newcommand{\Be}{\begin{equation}}
\newcommand{\ee}{\end{equation}}

\newcommand{\half}{\frac{1}{2}}

\begin{document}

\preprint{NORDITA-2003-52 HE}

\title{A quantum BRST anti-BRST approach to classical integrable systems}
\author{Michael Chesterman}
 \email{Michael.Chesterman@nbi.dk}
\affiliation{Nordita, Blegdamsvej 17, Copenhagen, DK-2100,
Denmark.}
\author{Marcelo B. Silka}
 \email{silka@nbi.dk}
\affiliation{Niels Bohr Institite, Blegdamsvej 17, Copenhagen,
DK-2100, Denmark.}

\date{August 2003}

\begin{abstract}
We reformulate the conditions of Liouville integrability in the
language of Gozzi \textit{et al.}'s quantum BRST anti-BRST
description of classical mechanics. The Das-Okubo geometrical Lax
equation is particularly suited to this approach. We find that the
Lax pair and inverse scattering wavefunction appear naturally in
certain sectors of the quantum theory.
\end{abstract}

\pacs{02.30.Ik}
\maketitle

\section{Introduction}

We use the quantum mechanical approach to classical mechanics,
developed by Gozzi, Reuter and Thacker
\cite{Gozzi:1988ge,Gozzi:1989bf, Gozzi:1992di} and generalized by
Marnelius \cite{Marnelius:2000vc} to a symplectic supermanifold,
in order to study Liouville integrability. While it is novel in
itself to obtain classical mechanics from a quantum system, our
real interest in this approach is with the natural way in which it
marries Hamiltonian mechanics with differential geometry.

Integrable systems are of considerable interest to physicists,
mostly because their equations of motion possess soliton
solutions. There are various equivalent ways to establish whether
a system is integrable, for example the construction of a Lax
equation or the zero curvature formalism. In this article, we
study two well-established methods, both of which require the
existence of a bi-Hamiltonian structure. Both of these methods
were originally constructed using standard Hamiltonian mechanics.
We aim to show here the advantages of using the quantum BRST
description of classical mechanics.

The paper is structured as follows. In section \ref{sec:setup} we
set up the bi-Hamiltonian system.

In section \ref{sec:integrable_Magri}, we use the technique of
Magri \cite{Magri:1978gn}, to directly derive integrability from
the existence of a bi-Hamiltonian structure. A further condition
is also needed that the two Poisson brackets obey a certain
compatibility condition. The quantum BRST approach is nice here
because geometrical constructs, such as the Hamiltonian flow
vector field, actually belong to the phase space manifold. This
was first studied in an interesting paper by Calian
\cite{Calian:2000es}, though we follow a slightly different route.

In section \ref{sec:integrable_Das}, we use the approach of Das
and Okubo\cite{Okubo:1988nh} to derive a Lax equation from the
bi-Hamiltonian structure. The system is found to be integrable
when the Nijenhuis tensor, which depends on the Lax operator,
vanishes. Here the mechanics of Gozzi \textit{et al.} really comes
into its own. In particular, the Lax equation actually \textit{is}
the Hamiltonian equation of motion of the Lax tensor, and the
associated inverse scattering wavefunction appears at ghost number
one. We create more general Lax Pairs, by deforming the ghost
co-ordinates, while maintaining the BRST anti-BRST symmetry
algebra of the theory. We exploit the fact that the Lax equation
depends on the existence of this algebra and not on the brackets
of the co-ordinates themselves.

Finally, we note some interesting earlier work
\cite{Okubo:1991ub,Das:1990cp,Okubo:1991xu}, which has some
similarity in approach to this article.

\section{Preliminaries}\label{sec:setup}
\subsection{The symplectic manifold}

For simplicity we will only consider bosonic systems,
but it is straightforward to extend this paper to fermionic
systems. We begin with a symplectic manifold $\M$ with phase-space
co-ordinates $\phi^a$ for $a =1 ,\ldots, 2N$, which is endowed
with two symplectic forms
\begin{eqnarray}
\om_r = \half (\om_r)_{ab}(\phi) d\phi^a \wedge d\phi^b, && r =0,
1,
\end{eqnarray}
which are non-degenerate and closed
\begin{eqnarray}\label{eq:om_non_deg_closed}
\det \om_r \neq 0, && d\om_r = 0.
\end{eqnarray}
Each two-form $(\om_r)_{ab}$ has an inverse $\om_r^{ab}$ defined
by
\begin{eqnarray}
(\om_r)_{ab}(\om_r)^{bc} = \de_a^c,
\end{eqnarray}
with which is associated the Poisson bracket
\begin{eqnarray}\label{PB2n}
\{ f(\phi),g(\phi)\}_{r}=(\pa_a f) \om^{ab}_{r}(\pa_b g), &&
{r}=0,1.
\end{eqnarray}
The bracket obeys the Jacobi identity due to $\om_r$ being closed
\eqref{eq:om_non_deg_closed}, and is antisymmetric under
interchange of $f$ and $g$, since $\om_r^{ab}=-\om_r^{ba}$.

We associate a Hamiltonian $H_1(\phi)$ with symplectic form
$\om_1$ and $H_2(\phi)$ with $\om_0$. There exists a
bi-Hamiltonian structure if each bracket plus Hamiltonian pairing
yields the same equations of motion
\begin{equation}\label{eq:original_bi_ham_eqn}
\dot{\phi}^a = \om_1^{ab}\pa_b H_1 = \om_0^{ab}\pa_b H_2.
\end{equation}

\subsection{The BRST anti-BRST quantum description of classical mechanics}

Now, we look at the same system, but using the BRST quantum
mechanical approach
\cite{Gozzi:1988ge,Gozzi:1989bf,Gozzi:1992di,Marnelius:2000vc}. We
extend the phase space with bosonic variables $\la_a$ and
fermionic ghosts $C^a$, $\PP_a$. From now on we work with quantum
operators, which are denoted by a hat, and set $\hbar=1$.

The non-vanishing, graded commutators for the extended phase-space
co-ordinates are defined by
\begin{eqnarray}\label{PB8n}
[ \phih^a,\lah_b ]=\delta^a_b, && [ \Ch^a,\PPh_b ]=\delta^a_b,
\end{eqnarray}
where the usual graded quantum bracket $[\cdot,\cdot]$ is
\begin{equation}
[\Ah , \Bh ]= \Ah \Bh - (-)^{\eps_A \eps_B} \Bh \Ah,
\end{equation}
such that $\eps_A$ is the Grassmann parity of $\Ah$. So here
$\eps_\phi=\eps_\la =0$ and $\eps_C = \eps_\PP=1$. The bracket
also obeys the Jacobi identity
\begin{eqnarray}
[[\Ah ,\Bh ],\Ch ] &+ (-)^{\eps_A(\eps_B+\eps_C)}[[\Bh ,\Ch ],\Ah ]\nonumber\\
&+ (-)^{\eps_C(\eps_A + \eps_B)} [[\Ch ,\Ah ],\Bh ]=0.
\label{eq:jacobi}
\end{eqnarray}
Note in particular that $\phih^a$ commute with each other, so
there are no ordering ambiguities for operators of the form
$A(\phih)$, which is what we want in order to describe classical
mechanics. Equation \eqref{PB8n} follows directly in the
Schr\"odinger representation from the operator definitions
\begin{eqnarray}
\phih^a \equiv \phi^a, &&\lah_a \equiv -\pa_a \\
\Ch^a \equiv C^a, && \PPh_a \equiv \frac{\pa}{\pa C^a}.
\end{eqnarray}
We define BRST and anti-BRST charges as in \cite{Marnelius:2000vc}
\begin{eqnarray}
\Qh=\Ch^a \lah_a, && \Qbh_{r} = \PPh_a \om_{r}^{ab} \lah_b +
\frac{1}{2}(\pa_c \om_{r}^{ab})\Ch^c \PPh_a \PPh_b
\end{eqnarray}
which obey
\begin{eqnarray}
[\Qh,\Qh]=  [\Qbh_{r} ,\Qh]= [\Qbh_{r},\Qbh_{r} ]=0,
\end{eqnarray}
using equation \eqref{eq:om_non_deg_closed} and that $\om_r$ is
antisymmetric.

The time evolution for any operator $\Fh$ is given by
\begin{equation}\label{eq:Fdot}
\dot{\Fh} = [\Fh , \Hh_{eff}],
\end{equation}
where
\begin{equation}\label{Heff}
\Hh_{eff}=-[\Qh, [\Qbh_{1} ,\Hh_1]] = -[\Qh, [\Qbh_{0} ,\Hh_2]]
\end{equation}
is the effective Hamiltonian\cite{Gozzi:1989bf,Marnelius:2000vc}.
Note that the bracket \eqref{PB8n} is flat and $\Hh_{eff}$
contains both the information of the original Hamiltonian and of
the original curved Poisson bracket \eqref{PB2n}. The above two
definitions of $\Hh_{eff}$ are equal as a result of the
bi-Hamiltonian equation \eqref{eq:original_bi_ham_eqn}. In
components, using $\om_1$ and $H_1$ for example,
\begin{equation}\label{eq:H_eff_cpts}
\Hh_{eff} =\lah_a \omh_1^{ac}(\pa_c \Hh_1) - \Ch^a \PPh_b
\pa_a(\omh_1^{bc}\pa_c \Hh_1).
\end{equation}
While $\dot{\phih}^a = [\phih^a , \Hh_{eff}]$ is the same as in
equation \eqref{eq:original_bi_ham_eqn}, the ghosts $\Ch^a$ and
$\PPh_a$ obey
\begin{eqnarray}
\dot{\Ch}^b = \Ch^a {\Uh_a}^b && \dot{\PPh}_a = - \PPh_b
{\Uh_a}^b,
\end{eqnarray}
where ${\Uh_a}^b = \pa_a( \omh_1^{bc}\pa_c \Hh_1)$. The above
equations of motion for $\Ch^b$ and $\PPh_a$ are exactly the same
as for the 1-form $d\phi^b$ and the vector $\pa/\pa\phi^a$
respectively. Thus in general we can study the time evolution of a
generic $(p,q)$ tensor
\begin{equation}\label{eq:tensor_T}
\Th= {T_{a_1 ... a_q}}^{b_1 ... b_p}(\phih) \Ch^{a_1}...\Ch^{a_q}
\PPh_{b_1}...\PPh_{b_p},
\end{equation}
which is separately antisymmetric in its upper and lower indices
since the ghosts are fermionic. We interpret $\Ch^a$ as $d\phi^a$
and $\PPh_a$ as $\pa/\pa\phi^a$, and the ghost products
\begin{eqnarray}
\Ch^{a_1} \ldots \Ch^{a_q} \equiv d\phi^{a_1} \wedge \ldots \wedge
d\phi^{a_q}\\
\PPh_{b_1} \ldots \PPh_{b_p} \equiv \frac{\pa}{\pa\phi^{b_1}}
\wedge \ldots \wedge \frac{\pa}{\pa\phi^{b_p}},
\end{eqnarray}
where we extend the definition of the wedge product to vectors in
the same way as with forms i.e. as an antisymmetric direct
product. Whereas in ordinary Hamiltonian mechanics, we are limited
to writing the time evolution of functions, the extension to
$(p,q)$ antisymmetric tensors will be very useful to us in section
\ref{sec:integrable_Das}.

As well as $\Qh$ and $\Qh_r$, other operators which commute with
$\Hh_{eff}$ are the antisymmetric tensors
\begin{eqnarray}\label{eq:K_defn}
\Kh_r = \half (\omh_r)_{ab} \Ch^a \Ch^b, && \Kbh_r = \half
\omh_r^{ab} \PPh_a \PPh_b,
\end{eqnarray}
and the ghost number operator $\Qh_g = \Ch^a \PPh_a$. Their
algebra, which is the symmetry algebra of this system, is given in
equation \eqref{alg}. We derive that $\Kh_r$ and $\Kbh_r$ commute
with $\Hh_{eff}$ later in equation \eqref{eq:K_H_eff}.

A geometrical interpretation of the bi-Hamiltonian equation
\eqref{eq:original_bi_ham_eqn} is that the vector field
\begin{equation}\label{eq:BiHam_eqn}
X = [ \Hh_2 , \Qbh_0 ] =[ \Hh_1, \Qbh_1 ].
\end{equation}
describing the Hamiltonian flow of the system, has two equivalent
Hamiltonian descriptions. Of course the vector field is a special
case of the tensor \eqref{eq:tensor_T}. We see that in components,
$[\Hh_1 , \Qbh_1 ]$ = $\PPh_a \om_1^{ab} \pa_b \Hh_1 = \PPh_a
\dot{\phih}^a$ and similarly for $[ \Hh_2 , \Qbh_0]$, where we
recall that $\PPh_a \equiv \pa/\pa \phi^a$.

We will often need to write the original Poisson bracket
\eqref{PB2n} in terms of the bracket of the quantum theory. There
are various equivalent expressions. The ones that we use are
stated below, and some others are listed in the appendix.

Firstly, note that the expression \cite{Gozzi:1989bf} for a vector
field $X_{f_r}=[ f(\phih), \Qbh_r ]$ acting on a function
$g(\phi)$ is
\begin{equation}\label{eq:X_g}
X_{f_r} g = [X_{f_r}, [\Qh , g ]],
\end{equation}
which is of course the same as the Poisson bracket $\{f,g\}_r$.
This leads to two equivalent expressions for a bracket, depending
on whether we calculate $X_{f_r} g$ or $-X_{g_r} f$
\begin{equation}\label{eq:PB_Marnelius_variation}
\{f(\phi) , g(\phi) \}_r = [[f , \Qbh_r] , [\Qh , g]] = -[[f ,
\Qh] , [\Qbh_r , g]].
\end{equation}
These expressions are closely related to Marnelius' bracket
\eqref{eq:Marn_bracket}, which is an extension of the above to
allow general functions $f$ and $g$ of the full phase-space.

Finally, we note that many more differential geometry operations
than listed here can be expressed \cite{Gozzi:1989bf,Gozzi:1992di}
as brackets between quantum operators, or as operators acting on
the Hilbert space, which is isomorphic to the space of p-forms.


\section{Direct proof of integrability of a bi-Hamiltonian system}\label{sec:integrable_Magri}

The requirements of Liouville integrability are that there exist
an infinite set of conserved charges, which commute with each
other under the Poisson brackets \eqref{PB2n}. In
\cite{Magri:1978gn}, Magri proved that all bi-Hamiltonian systems
are integrable, given that the two Poisson brackets obey a certain
compatibility condition described below. We give the same proof,
but using Gozzi $\textit{et al.}$'s quantum BRST description of
classical mechanics.

Given the bi-Hamiltonian equation \eqref{eq:BiHam_eqn}, one can
show that there always exists a third operator $\Hh_3(\phi)$ such
that the vector field $Y$ has two equivalent descriptions
\begin{equation}\label{eq:H_3}
Y =[ \Hh_3, \Qbh_0 ] =[ \Hh_2, \Qbh_1 ],
\end{equation}
so long as the Poisson brackets $\{,\}_0$ and $\{,\}_1$ are
compatible, i.e. that the bracket $\{,\}_0 + \{,\}_1$ is
non-degenerate and obeys the Jacobi identity. In order to prove
existence of $H_3$, we must show that the vector field $Y$ is
Hamiltonian with respect to the bracket $\{,\}_0$. The Jacobi
identity for $H_2$, $f$, $g$ and Poisson bracket $\{,\}_0 +
\{,\}_1$ yields
\begin{equation}\label{eq:compatibility}
\{ H_2,\{ f,g\}_1\}_0 + \{ f,\{ g,H_2\}_0\}_1 + \{ g,\{
H_2,f\}_0\}_1 + \, _{(0 \leftrightarrow 1)}=0,
\end{equation}
where $f$ and $g$ are arbitrary functions of $\phi^a$. Since $X$
in equation \eqref{eq:BiHam_eqn} is Hamiltonian with respect to
$\om_1$ by definition, using the Jacobi identity we find that
\begin{equation} \label{eq:X_f_g_1}
X\{f,g\}_1 - \{Xf,g\}_1 - \{f,Xg\}_{1}=0,
\end{equation}
thus from \eqref{eq:compatibility} and \eqref{eq:X_f_g_1}
\begin{equation}\label{eq:X'_Ham_om_0}
Y \{f,g\}_0 - \{Yf,g\}_0 - \{f,Yg\}_0 = 0,
\end{equation}
where $Y = [H_2 ,\Qb_1]$. Equation \eqref{eq:X'_Ham_om_0} is a
necessary and sufficient condition that $Y$ is Hamiltonian with
respect to $\om_0$. We could have completed the above stages using
the Marnelius style brackets \eqref{eq:PB_Marnelius_variation} and
the definition of a vector acting on a function \eqref{eq:X_g}. In
the end it's simpler as above.

From \eqref{eq:H_3}, by iteration there exist $\Hh_m$, $m \geq 1$
such that
\begin{equation}\label{eq:Q_bar_H_m}
[ \Hh_{m+1}, \Qbh_0] = [ \Hh_m, \Qbh_1].
\end{equation}

Using the Marnelius style Poisson bracket
\eqref{eq:PB_Marnelius_variation} and equation
\eqref{eq:Q_bar_H_m},
\begin{eqnarray} \label{eq:H_n_H_m_bracket1}
\{H_n, H_{m+1}\}_{0} = [[\Hh_n,\Qh],[\Qbh_0,\Hh_{m+1}]]
=[[\Hh_n,\Qh],[\Qbh_1,\Hh_{m}]]=\{H_n, H_{m}\}_{1},
\end{eqnarray}
and similarly
\begin{equation} \label{eq:H_n_H_m_bracket2}
\{H_n, H_{m}\}_{1} = -[[\Hh_n,\Qbh_1],[\Qh_,\Hh_{m}]] =
-[[\Hh_{n-1},\Qbh_0],[\Qh_,\Hh_{m}]] = \{H_{n-1},H_m\}_{0}.
\end{equation}
Now $\{H_{n} , H_n\}_{r}$ is zero since the Poisson bracket is
antisymmetric and $\{H_n , H_{n+1}\}_{r}=0$ from either
\eqref{eq:H_n_H_m_bracket1} or \eqref{eq:H_n_H_m_bracket2}.
Applying recursion relations \eqref{eq:H_n_H_m_bracket1} and
\eqref{eq:H_n_H_m_bracket2}, any commutator $\{H_n , H_{m+1}\}$
can be written either in the form $\{H_{p} , H_p\}_{r}$ or $\{H_p
, H_{p+1}\}_{r}$ for some integer $p$, and hence vanishes.
Therefore
\begin{equation}
\{H_m , H_n\}_r = 0.
\end{equation}
Since all $H_m$ commute with $H_1$ and $H_0$, they are all
constants of motion.

\section{The Das-Okubo geometrical Lax equation}\label{sec:integrable_Das}

In \cite{Okubo:1988nh}, Das and Okubo found an alternative way of
deriving integrability of a bi-Hamiltonian system, by providing a
recipe for a Lax pair. This gives the Lax pair method a
geometrical interpretation, in terms of the bi-Hamiltonian
structure, and relates two seemingly unconnected constructions. As
explained in the introduction, the quantum BRST approach to
classical mechanics is particularly fruitful here.

As in the previous section, the requirements of Liouville
integrability are that there exist an infinite set of conserved
charges, which commute with each other under the Poisson brackets
\eqref{PB2n}.

Using the expression for $\Hh_{eff}$ in equation \eqref{Heff}, the
Jacobi identity \eqref{eq:jacobi} and the BRST anti-BRST algebra
of equation \eqref{alg}, we find that
\begin{eqnarray}\label{eq:K_H_eff}
[\Kh_r , \Hh_{eff}] = [\Kbh_r,\Hh_{eff}] =0
\end{eqnarray}
where recall that $\Kh_r \equiv \om_r$ and $\Kbh_r \equiv
\om_r^{-1}$ as in equation \eqref{eq:K_defn}. Thus, the tensors
$\Kh_r$ and $\Kbh_r$ are invariant under Hamiltonian flow. Note
that this does not mean that $(\om_r)_{ab}$ is constant, rather
that its time evolution is cancelled by the time evolution of the
ghosts $\Ch^a$, and similarly for $\om_r^{ab}$ and $\PPh_a$.
Recall that $\Ch^a$ is identified with $d\phi^a$, and $\PPh_a$
with $\pa/\pa\phi^a$.

We define the (1,1) tensor
\begin{equation}
\Sh=[\Kh_1,\Kbh_0]= \Ch^a {S_a}^{b}(\phih) \PPh_b \equiv
{S_a}^b(\phi)\,d\phi^a \otimes \frac{\pa}{\pa\phi^b},
\label{eq:Sh}
\end{equation}
where ${S_a}^{b} =(\omh_{0})_{ac}\omh_1^{cb}$, as a candidate for
a Lax operator. From equation \eqref{eq:K_H_eff} and the Jacobi
identity \eqref{eq:jacobi}, $\Sh$ is also invariant under
Hamiltonian flow
\begin{equation}\label{eq:S_H_eff}
\dot{\Sh}=[\Sh, \Hh_{eff}]=0.
\end{equation}
Utilizing the expression for $\Hh_{eff}$ in equation
\eqref{eq:H_eff_cpts}, the above equation reads
\begin{equation}\label{eq:Lax_equation_cpts}
\Ch^a (\dot{{S_a}^b} - {[S,U]_a}^b)\PPh_b=0,
\end{equation}
where
\begin{equation}\label{eq:U}
\Uh=\Ch^a \pa_a(\om_1^{bc}\pa_c H_1)\PPh_b = \Ch^a
\pa_a(\om_0^{bc}\pa_c H_2)\PPh_b= \Ch^a {U_a}^b(\phi)\PPh_b,
\end{equation}
and we used that
\begin{equation}
[\Sh , \Uh ] = \Ch^a {[S,U]_a}^b \PPh_b,
\end{equation}
where ${[S,U]_a}^b= {S_a}^c {U_c}^b - {U_a}^c {S_c}^b$. This is
indeed the same form as the Lax equation with the tensors $\Sh$
and $\Uh$ as the Lax Pair.

So, the Lax equation \eqref{eq:Lax_equation_cpts} actually is
simply the quantum Hamiltonian equation of motion for the Lax
operator $\Sh$, although it also implies the classical equations
of motion for $\phi^a$. It seems slightly mysterious that the Lax
equation on the one hand is interpreted as a quantum mechanical
equation (with $\hbar=1$), as in the inverse scattering method
\cite{Das:2001ad}, and on the other hand implies the classical
equations of motion. However, this is a natural feature of our
quantum BRST approach to classical integrable systems.

We construct the conserved charges, using
\begin{eqnarray}\label{eq:Sn_H_eff}
[\Sh^n,\Hh_{eff}]=0, && n =\pm 1, \pm 2, \dots,
\end{eqnarray}
which follows from equation \eqref{eq:S_H_eff} and the Jacobi
identity \eqref{eq:jacobi}, where $\Sh^{-1}$ is defined by
$<\psi_a|\Sh^{-1} \Sh|\psi^b> = \de_a^b$. Taking the quantum trace
of equation \eqref{eq:Sn_H_eff} with respect to $<\psi_a|$ and
$|\psi^a>$, which are defined in equation \eqref{eq:states}, we
find there are conserved charges
\begin{eqnarray}
I_n(\phi) =\frac{1}{n} \sum_a <\psi_a|\Sh^n|\psi^a> && n =\pm 1,
\pm 2, \dots
\end{eqnarray}
\begin{equation}
\dot{I}_n = 0.
\end{equation}
Note that the inner product is taken only over ghosts and not over
$\phi^a$, so $<\psi_a|\Sh|\psi^b> = {S_a}^b(\phi)$ for example,
and $<\psi_a|\Sh^n|\psi^b> = {(S^n)_a}^b$. In general not all
$I_n$ will be functionally independent for finite $N$, since $\Sh$
is a $2N\times 2N$ matrix \cite{Okubo:1988nh}. For example, all
$I_n$ with $n \geq 2N+1$ can be expressed as polynomials of $I_1,
\ldots, I_{2N}$. If there are $N$ $I_n$'s which are functionally
independent, we have the correct number for integrability. Of
course for most interesting examples, the $a$ in $\phi^a$ is a
continuous parameter, and $N$ is therefore infinite.

Finally, we require that the charges commute
\begin{equation}
\{I_n,I_m\}_{0}=\{I_n,I_m\}_1=0,
\end{equation}
which we have learnt in section \ref{sec:integrable_Magri} amounts
to asking that equation \eqref{eq:Q_bar_H_m} be satisfied, with
$H_n$ replaced by $I_n$. Using \eqref{eq:Sh} and \eqref{alg}, and
reversing the roles of $\om_0$ and $\om_1$ we rewrite equation
\eqref{eq:Q_bar_H_m} as
\begin{equation}\label{eq:I_n_bi_Ham}
[\Sh , [\Qh ,I_{n}]] - [\Qh, I_{n+1}] = 0.
\end{equation}
We define the operator
\begin{equation}\label{eq:N}
\Nh=2 [\Qh,\Sh] \Sh  = \Ch^a \Ch^b {N_{ab}}^c(\phih)\PPh_c +
\ldots
\end{equation}
where the ellipses refer to terms with different ghost
contributions, and
\begin{equation}\label{eq:N_ab_c}
{N_{ab}}^c ={S_a}^d \pa_d {S_b}^c - {S_b}^d \pa_d {S_a}^c - (\pa_a
{S_b}^d) {S_d}^{c}  + (\pa_b {S_a}^d) {S_d}^c
\end{equation}
is the Nijenhuis tensor. Writing equation \eqref{eq:I_n_bi_Ham} in
components, we find that
\begin{eqnarray}
{S_a}^b \pa_b I_{n} - \pa_a I_{n+1} = {N_{ab}}^c
{{S^{(n-1)}}_c}^b,
\end{eqnarray}
therefore the condition of Liouville integrability amounts to
requiring ${N_{ab}}^c=0$.

From the definition of $\Nh$ \eqref{eq:N}, the Jacobi identity
\eqref{eq:jacobi}, and using that $\Sh$ commutes with $\Hh_{eff}$
\eqref{eq:S_H_eff}, as does $\Qh$ because $\Hh_{eff}$ is
$\Qh$-exact \eqref{Heff}, we find
\begin{equation}
\dot{\Nh}=[\Nh , \Hh_{eff}] =0.
\end{equation}
Therefore, ${N_{ab}}^c=<\psi_{ab}|\Nh|\psi^c>=0$, where $|\psi^c>$
and $|\psi_{ab}>$ are defined in the appendix, is consistent with
time evolution.

Another interesting point is that the ghost number one
wavefunction $\psi_1 = C^a f_a(\phi)$ is associated with the Lax
equation in the inverse scattering method \cite{Das:2001ad}. This
uses quantum mechanical techniques, in particular the (linear)
equation of motion for the wavefunction $\psi_1$, given by
\begin{equation}
\dot{\psi_1} = - \Uh \psi_1 = -\Ch^a {{U_a}^b}f_b,
\end{equation}
to solve the (non-linear) equations of motion for $\phi^a$. The
configuration of $\phi^a$ is encoded in $\psi_1$. Although so far
we have considered only the time evolution of operators as in
equation \eqref{eq:Fdot}, the equations of motion for the
classical system can also be described as a Schr\"odinger equation
for the ghost number 0 wavefunction $\psi_0(\phi)$
\cite{Gozzi:1992di,Marnelius:2000vc}.

Our recipe for constructing a Lax equation \eqref{eq:S_H_eff} is
essentially reliant only on the existence of the algebra between
$\Qh$, $\Qbh$, $\Kh_r$, $\Kbh_r$ in equation \eqref{alg}. It is
therefore natural to ask how one can deform the phase-space while
maintaining this algebra, in order to make new Lax pairs for the
same system. The equations of motion for $\phih^a$ are determined
only by the first term in $\Hh_{eff}$ in equation
\eqref{eq:H_eff_cpts}, because the second term is independent of
$\lah_a$. Since we wish to keep the same equations of motion for
$\phih^a$, we can only deform the second term in $\Hh_{eff}$. We
choose the following redefinition of the ghosts
\begin{eqnarray}
{\Ch'^{a}} = \Ch^b {A(\phi)_b}^a,  && {\PPh'_a} =
{A^{-1}(\phi)_a}^b \PPh_b ,
\end{eqnarray}
where $A$ is chosen such that $\det A \neq 0$, and that the BRST
anti-BRST algebra (\ref{alg}) remains the same. From definitions
(\ref{eq:Sh}) and (\ref{eq:U}), for $\Sh$ and $\Uh$, we have
\begin{eqnarray}\label{eq:sim_transform_S_U}
\Sh' = \Ch^a {(A S A^{-1})_a}^b \PPh_a, && \Uh' = \Ch^a {(A U
A^{-1})_a}^b \PPh_b.
\end{eqnarray}
Given that the BRST algebra \eqref{alg} has been maintained,
$\Sh'$ obeys equation \eqref{eq:S_H_eff}, and $\Sh'$, $\Uh'$ obey
the Lax equation \eqref{eq:Lax_equation_cpts}, which implies the
equations of motion for $\phih^a$ \eqref{eq:original_bi_ham_eqn}.

In the case where $\pa_c{A_a}^b=0$, it is clear that the quantum
brackets between phase-space co-ordinates in equation \eqref{PB8n}
are unaltered, in particular $[ {\Ch'^a} , \lah_b ] =0$, therefore
the algebra \eqref{alg} is also unchanged. For non-constant $A$,
the co-ordinate algebra is changed, for example $[ \Ch'^a  ,
\lah_b] \neq 0$. However the algebra \eqref{alg} is maintained if
$d\om_r '=0$, or in terms of $\Kh_r'$
\begin{equation}
[\Qh' , \Kh_r'] = \Ch^a \Ch^b \Ch^c {A_a}^f \pa_f ({A_b}^g
(\om_r)_{gh}{A_c}^h )=0.
\end{equation}
The above equation is satisfied iff
\begin{equation}
{A_a}^b = \frac{\pa \phi^b}{\pa \phi'^{a}},
\end{equation}
for some smooth invertible function $\phi'^a(\phi)$. In other
words, the matrix ${A_a}^b(\phi)$ must represent a co-ordinate
transformation of the symplectic manifold $\M$.

It would be interesting to investigate whether any Lax pair can be
written in the form of $\Sh'$ and $\Uh'$ in equation
\eqref{eq:sim_transform_S_U}.

\section{Conclusions} \label{sec:conclusions}
We have seen how the quantum BRST description of classical
mechanics has advantages over standard Hamiltonian mechanics in
the study of integrability. This is particularly apparent with the
derivation of the Das-Okubo geometrical Lax equation in section
\ref{sec:integrable_Das}. What's more, the mixture of quantum and
classical aspects of the Lax equation arise naturally in this
approach. Further investigation seems warranted.

\begin{acknowledgments}
We wish to thank Prof. Poul H. Damgaard for useful discussions.
M.B.S. thanks the IBCCF/UFRJ, where part of this work was written,
for their kind hospitality. M.C. was funded by a Marie Curie
training site fellowship.
\end{acknowledgments}

\begin{appendix}
\section{Useful identities}
The BRST anti-BRST symmetry algebra is given by
\begin{equation}\label{alg}
\begin{array}{c}
[\Qh,\Qh]=  [\Qbh_{r} ,\Qh]= [\Qbh_{r},\Qbh_{r} ]=0,\\
\begin{array}{ccc}
{[}\Qh_g, \Kh_{r} ] = \Kh_{r},& [\Qh_g,\Kbh_{r}]=-\Kbh_{r}, &
[\Kh_{r},\Kbh_{r} ] =\Qh_g ,
\end{array}\\
\begin{array}{cc}
{[}\Kh_{r},\Qh ] =0, & [\Kh_{r},\Qbh_{r} ] = \Qh ,\\
{[}\Kbh_{r} ,\Qh ] =\Qbh_{r}, & [\Kbh_{r} ,\Qbh_{r} ] = 0,\\
{[}\Qh_g,\Qh] = \Qh ,& [\Qh_g,\Qbh_{r} ] =-\Qbh_{r}.
\end{array}
\end{array}
\end{equation}

The Marnelius bracket\cite{Marnelius:2000vc} is defined as
\begin{equation}\label{eq:Marn_bracket}
\{ f,g\}_{\Qb_r}\equiv \frac{1}{2}\bigg ( [ [ \fh ,\Qbh_{r}] , [
\Qh , \gh]] - [ [\fh,\Qh] ,[ \Qb_{r} ,\gh]]\bigg ).
\end{equation}
Another interesting expression for the Poisson bracket is
\begin{equation}
\{ f,g\}_{r}=-[ X_{f_r},[ \Kh_{r},X_{g_r}] ],
\end{equation}
which is a bracket between two vector fields $X_{f_r}=[\fh ,\Qbh_r
]$ and $X_{g_r}=[\gh ,\Qbh_r ]$.

States of specific ghost number are constructed by applying the
$\Ch^a$ or $\PPh_a$ operators to the "ground" states
\begin{eqnarray}
|\PP=0> \equiv 1, && |C=0> \equiv \prod_{a}{C^a},
\end{eqnarray}
where the inner product $<C=0|\PP=0>=1$. For example, we define
the states
\begin{eqnarray}\label{eq:states}
|\psi^a> \equiv \Ch^a|\PP=0>, & |\psi_a> \equiv \PPh_a |C=0>, &
|\psi_{ab}> \equiv \PPh_a\PPh_b |C=0>.
\end{eqnarray}
\end{appendix}

\bibliography{integrable}
\end{document}